# Covariant formulation of Aharonov-Bohm electrodynamics and its application to coherent tunnelling

GIOVANNI MODANESE

*Free University of Bolzano, Faculty of Science and Technology*
*P.za Università 5, Bolzano, Italy*
*Giovanni.Modanese@unibz.it*

The extended electrodynamic theory introduced by Aharonov and Bohm (after an earlier attempt by Ohmura) and recently developed by Van Vlaenderen and Waser, Hively and Giakos, can be re-written and solved in a simple and effective way in the standard covariant 4D formalism. This displays more clearly some of its features. The theory allows a very interesting consistent generalization of the Maxwell equations. In particular, the generalized field equations are compatible with sources (classical, or more likely of quantum nature) for which the continuity/conservation equation $\partial_\mu j^\mu=0$ is not valid everywhere, or is valid only as an average above a certain scale. And yet, remarkably, in the end the observable $F_{\mu\nu}$ field is still generated by a conserved effective source which we denote as $(j^\nu+i^\nu)$, being $i^\nu$ a suitable non-local function of $j^\nu$. This implies that any microscopic violation of the charge continuity condition is "censored" at the macroscopic level, although it has real consequences, because it generates a non-Maxwellian component of the field. We consider possible applications of this formalism to condensed-matter systems with macroscopic quantum tunneling. The extended electrodynamics can also be coupled to fractional quantum systems.

*Keywords*: Maxwell equations; local charge conservation; Josephson junctions

## 1. Introduction

The Maxwell equations play a crucial role in physics and engineering. Their mathematical properties are a beautiful lesson of vector analysis and their Lorentz invariance is a strong precursor of Special Relativity. The quantization of the Maxwell field leads to Quantum Electrodynamics, one of the most successful and accurate physical theories, and the extension of the gauge principle is at the basis of modern theories of elementary particles. So there appears to be little room for improvement of the Maxwell theory. From the experimental point of view, the literature reports a few claims of detection of electromagnetic waves with longitudinal components [1,2], which are not compatible with the Maxwell equations.

An extension of Maxwell theory which is compatible with additional degrees of freedom in electromagnetic waves in vacuum is the Aharonov-Bohm electrodynamics [3-7]. The Aharonov-Bohm Lagrangian comprises an additional term proportional to $(\partial_\mu A^\mu)^2$ and has only a reduced gauge invariance; the Lorentz gauge is not applicable to this theory and no quantized versions have been proposed. So it cannot be regarded as a candidate replacement of Maxwell theory at a fundamental particle level, but possibly as adequate to the description of peculiar situations involving electromagnetic waves which are not purely transverse, but comprise a scalar component *S*.

The Aharonov-Bohm equations, however, set a clear limit to their own application when one considers the wave equation with sources. As already remarked in [4], this equation implies that a scalar electromagnetic field *S* can only be generated by a source which does not respect the local charge conservation condition $\partial_\mu J^\mu=0$. Hively and Giakos [7] discuss whether this could happen due to charge fluctuations at a microscopic level. Van Vlaenderen [5] hypothesized that macroscopic quantum tunnelling could imply that charge is not locally conserved (although in fact this can only happen if the conserved current of the Schrödinger equation or its multi-particle generalization is not applicable - compare below, Sect. 3).

Other possible applications concern situations in which the occurrence of a quantum anomaly or a space-time singularity causes a tunneling process between two vacuum states [8].

The new covariant formulation of Aharonov-Bohm electrodynamics will be discussed in Sect. 2, and possible applications in Sect. 3. Sect. 4 contains a brief conclusion and outlook.

## 2. Lagrangian formalism and field equations

Let us follow the standard covariant formalism as for instance in [9], with Heaviside units and $c=1$. In these units the Maxwell equations are simply written

$$\text{div}\mathbf{E} = \rho$$
$$\text{curl}\mathbf{E} = -\frac{\partial \mathbf{B}}{\partial t} \quad (1)$$
$$\text{div}\mathbf{B} = 0$$
$$\text{curl}\mathbf{B} = \mathbf{j} + \frac{\partial \mathbf{E}}{\partial t}$$

The electromagnetic field tensor $F^{\mu\nu}$, whose components are the electric field $\mathbf{E}$ and magnetic field $\mathbf{B}$, is defined as $F_{\mu\nu} = \partial_\mu A_\nu - \partial_\nu A_\mu$. The four-potential is $A^\mu = (V, \mathbf{A})$. The four-current is $j^\mu = (\rho, \mathbf{j})$ and the "continuity" equation, or local charge conservation equation is

$$\frac{\partial \rho}{\partial t} + \text{div}\mathbf{j} = 0 \Leftrightarrow \partial_\mu j^\mu = 0 \quad (2)$$

The Maxwell equations (1) are written, in terms of $F^{\mu\nu}$ and $j^\mu$, as

$$\partial_\mu F^{\mu\nu} = j^\nu$$
$$\frac{1}{2} \partial^\rho \varepsilon_{\rho\sigma\mu\nu} F^{\mu\nu} = 0 \quad (3)$$

Note that since $F^{\mu\nu}$ is antisymmetric, eq. (3.a) can only be solved if $\partial_\nu j^\nu = 0$. The homogeneous equation (3.b) is a consequence of the definition of $F^{\mu\nu}$.

The field equation (3.a) can be derived from the Lagrangian

$$L = -\frac{1}{4} F_{\mu\nu} F^{\mu\nu} - j_\mu A^\mu + \frac{1}{2}\kappa\left[\left(\partial_\mu A^\mu\right)^2 - \partial_\mu A^\nu \partial_\nu A^\mu\right] \quad (4)$$

Here $L$ is seen as a function of the fundamental dynamical variable $A^\mu(x)$ and is the most general possible relativistic invariant Lagrangian constructed with a four-vector field. The term proportional to $\kappa$ can be written as a four-divergence and gives in fact no contribution to the action $\Omega = \int d^4 x L(x)$. When $\kappa = 0$ one speaks of the "minimal" Lagrangian. In a gauge transformation $A^\mu \to A^\mu + \partial^\mu \phi$ the variation of $L$ is

$$\Delta L = j_\mu \partial^\mu \phi \quad (5)$$

However, if the current is conserved, this is equivalent to

$$\Delta L = \partial^\mu(\phi j_\mu) \quad (\text{if } \partial_\mu j^\mu = 0) \quad (6)$$

and therefore $L$ is gauge-invariant up to a four-divergence.

Aharonov and Bohm [4] have proposed to generalize the electromagnetic Lagrangian by adding a term $\frac{1}{2}\gamma\left(\partial_\mu A^\mu\right)^2$ to the minimal $L$. (In their paper a parameter $\gamma = \lambda^{-1}$ is introduced.) This modified theory has also been studied by others, with various techniques [3,5,6,7]. Here we would like to further analyze it in the four-dimensional Lagrangian formalism. The new addition to the Lagrangian, not being a four-divergence, changes the field equations as follows:

$$L_{A.B.} = -\frac{1}{4} F_{\mu\nu} F^{\mu\nu} - j_\mu A^\mu + \frac{1}{2}\gamma\left(\partial_\mu A^\mu\right)^2 \to$$
$$\to \partial_\mu F^{\mu\nu} = j^\nu + \gamma \partial^\nu \left(\partial_\alpha A^\alpha\right) \quad (7)$$

This equation has been also derived by Woodside [10], with a different convention on the sign and constants. Under a gauge transformation, the Aharonov-Bohm Lagrangian changes as follows:

$$\Delta L_{A.-B.} = j_\mu(\partial^\mu \phi) + \frac{1}{2}\gamma(\partial^\alpha \partial_\alpha \phi)^2 + 2(\partial^\alpha A_\alpha)(\partial^\alpha \partial_\alpha \phi) \quad (8)$$

This means that the theory is not gauge-invariant anymore (even if $\partial_\mu j^\mu = 0$). It is only invariant under reduced gauge transformations, such that $\partial^\alpha \partial_\alpha \phi = 0$.

Note that since the general invariance is lost, we cannot impose the familiar Lorenz gauge, choosing a new four-potential $A'^\mu$ such that $\partial_\mu A'^\mu = 0$. Therefore we must regard the quantity $S = \partial_\alpha A^\alpha$ as a non-trivial dynamical variable, and we are going now to find out more about it. Take the derivative $\partial_\nu$ of the field eq. (7) and remember that $F^{\mu\nu}$ is antisymmetric. We obtain

$$\partial_\nu j^\nu = -\gamma \partial_\nu \partial^\nu (\partial^\alpha A_\alpha) = -\gamma \partial^2 S \quad (9)$$

where $\partial^2$ denotes the D'Alembert operator $\partial^2 = \partial_\alpha \partial^\alpha$. Note that, as expected, the current is not generally conserved. Of course, everything boils down to Maxwell equations with $S = 0$ and conserved current if $\gamma = 0$.

From (9) we obtain an expression for $S$:

$$S = \partial_\alpha A^\alpha = -\frac{1}{\gamma} \partial^{-2}\left(\partial_\nu j^\nu\right) \quad (10)$$

The well-known operator $\partial^{-2}$ is linear and non-local, as can be seen passing to four-momentum space, where it is represented by $k^{-2}$. Eq. (10) allows us to write an expression for the variable $S$, essentially integrating over the "source" $\partial_\nu j^\nu$:

$$S(x) = -\frac{1}{\gamma} \int d^4 k e^{-ikx} \frac{k_\nu \tilde{j}^\nu(k)}{k^2} \quad (11)$$

Going back to (10), let us now rename the summation index (ν→β), take again the derivative $\partial^\nu$ and multiply by λ. We obtain

$$\gamma \partial^\nu \left( \partial_\alpha A^\alpha \right) = -\partial^\nu \partial^{-2} \left( \partial_\beta j^\beta \right) \quad (12)$$

Therefore the generalized Maxwell equations (7) can be rewritten as follows:

$$\begin{cases} \partial_\mu F^{\mu\nu} = j^\nu + i^\nu \\ i^\nu = -\partial^\nu \partial^{-2} \left( \partial_\beta j^\beta \right) \end{cases} \quad (13)$$

Note that although these are derived from the Aharonov-Bohm Lagrangian $L_{A.B.}$, the parameter γ has disappeared. If the current $j^\mu$ is conserved, then the usual Maxwell equations are recovered. The new current component $i^\nu$ which now contributes, together with $j^\nu$, to generate the field $F^{\mu\nu}$, is such that the total current $\left( j^\nu + i^\nu \right)$ is always conserved, as can be checked in two ways: (1) by taking the derivative $\partial_\nu$ in (13.a); (2) by taking the derivative $\partial_\nu$ in (13.b), which yields consistently

$$\partial_\nu i^\nu = -\partial_\nu \partial^\nu \partial^{-2} \left( \partial_\beta j^\beta \right) = -\partial_\beta j^\beta \Rightarrow \partial_\nu \left( j^\nu + i^\nu \right) = 0 \quad (14)$$

Summarizing, we can say that the input of the generalized electrodynamic equations (13) is a four-current $j^\nu$ which is not necessarily conserved (computed, for instance, from an "anomalous" microscopic model, as discussed in the following); but the output is an electromagnetic field tensor $F^{\mu\nu}$ which has the usual properties, including that of being generated by a conserved current, namely $\left( j^\nu + i^\nu \right)$. It follows the important property that at the macroscopic level the current is always conserved, as far as it is possible to measure it through the field it generates. In other words, even though in this model the microscopic current $j^\nu$ can be not locally conserved, the observable current $\left( j^\nu + i^\nu \right)$ is always conserved. Since eq. (13.a) is linear, the field $F^{\mu\nu}$ is the sum of the fields generated by the currents $j^\nu$ and $i^\nu$. In general, the difference between the two currents is that even if the "primary" current $j^\nu$ is confined in a certain region of spacetime, the "secondary" current $i^\nu$ is not, because of the non-local expression which relates it to $j^\nu$.

Another surprising aspect of this generalized electrodynamics is the following. The new degree of freedom is the scalar quantity $S = \partial_\alpha A^\alpha$; in the traditional view this is a pure gauge mode and cannot contribute to $F^{\mu\nu}$; but here the dynamics is such that S affects the observable fields $F^{\mu\nu}$ through the secondary current $i^\nu$ which compensates for the local non-conservation of the primary microscopic current $j^\nu$.

For some static solutions giving configurations of $F^{\mu\nu}$ with planar and dipolar sources see [11].

## 3. A first possible application: locally non-conserved current in phenomena of coherent tunneling

We have seen that the new general equations admit solutions also when the microscopic current is not locally conserved, and yield then an "observable current" which is conserved. This reminds other situations typical of quantum mechanics, where one defines the theory in terms of microscopic quantities, like the wave function, which are not directly observable (while the usual macroscopically observable quantities may be not well-defined at the microscopic level).

We would like to explore the idea, originally proposed in [6] without any formal justification, that in phenomena of quantum tunneling the local conservation of the current might not be ensured. Also in quantum mechanics, however, it is possible to define a microscopic current which is locally conserved. For any solution Ψ of the Schrödinger equation one has

$$\rho = |\Psi|^2; \quad \mathbf{j} = \frac{-i\hbar}{2m} \left( \Psi^* \nabla \Psi - \Psi \nabla \Psi^* \right); \quad \frac{\partial \rho}{\partial t} + \nabla \cdot \mathbf{j} = 0 \quad (15)$$

In some tunneling devices which are operated with high precision, like for instance the tunneling effect microscope, this microscopic expression for the current has been accurately verified [12]. Analogous properties hold for the Ginzburg-Landau equation, which is a non-linear extension of the Schrödinger equation for the description of the macroscopic wave function of superconductors.

There are good reasons to believe, however, that in other less ideal situations it is too restrictive to assume the validity of an equation like (15). In condensed matter systems, macroscopic wave functions obey constrained equations and have therefore in general a non-locally conserved current [13].

In the second-quantization formalism, the current operator is conserved for free fields or in the presence of local interactions. This leaves the possibility of anomalous local non-conservation for certain state averages, or in the presence of non-local interactions.

Let us consider, for instance, a 1D tunneling process, in a stationary situation where a current flows across several barriers in series. Assume that $\frac{\partial \rho}{\partial t} = 0$ everywhere and therefore according to the continuity equation we should have $\frac{\partial(\rho v)}{\partial x} = 0$, i.e. $\rho v = const$. This means that the charged "fluid" must ideally adjust its velocity everywhere in inverse proportion to its density, so that the flux $\rho v$ is constant in each section of the material. At those places, deep inside the barriers, where ρ is very small, $v$ must be vary large. The Schrödinger or Ginzburg-Landau equations do not enforce any upper limit on $v$, but in reality we can expect some complications.

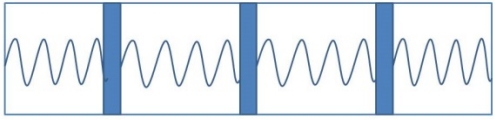

**Figure 1.** Macroscopic wave function of a supercurrent flowing through a 1D superconductor made of grains (like sintered YBCO). The inter-grain junctions conduct by Josephson tunneling or proximity effect. The amplitude is preserved in the tunneling.

Before discussing further the implications of the relation $\rho v = const$, we observe that a stationary tunneling flux is realized when a supercurrent flows across Josephson junctions in series (Fig. 1). This is a phenomenological model employed for the description of conduction in superconducting materials like YBCO or BSSCO, which have a granular structure and exhibit Josephson tunneling both of the intrinsic kind (between crystal layers with spacing of the order of 1 nm) and across the inter-grain junctions [15].

Now focus on the junctions (Fig. 2). Note that the amplitude of Ψ on the two sides of the junction must be the same, in order to allow for a complete transmission of the supercurrent. In an ideal case "a la Schrödinger", in order to keep $\rho v$ constant, the velocity $v$ of the pairs should increase very much in the center of the junction. It is known from experiments on the proximity effect that the critical current of a SNS junction decreases exponentially as exp(-$d$/$\xi_N$), where $\xi_N$ is a correlation length typical of the normal material (of the order of 1 nm, while $d$ is up to 100 nm). This is also confirmed by the non-local microscopic theory of Gorkov and De Gennes [14,16]. The Gorkov integral equation for superconductors is

$$\Delta(\mathbf{r}) = \int d^3\mathbf{r}' K(\mathbf{r},\mathbf{r}',T)\Delta(\mathbf{r}') \qquad (16)$$

where $\Delta(\mathbf{r})$ is the product of the pairs wavefunction and the effective electron interaction potential. The Ginzburg-Landau equation is a local approximation of this equation. The non-local Gorkov theory for the proximity effect also applies to other condensed matter systems, like semiconductors, where a wave function diffuses from one material into another one.

If we want to approximate this complex situation with an effective current density $j=\rho v$, the local conservation of $j$ in a stationary flow is only possible if $v$ increases exponentially exactly in parallel with the decrease of ρ. This would require an extreme superfluid behavior in a normal material, which is hardly compatible, in our opinion, with all the other interactions present.

Two possible alternatives are depicted in Fig. 2. The law $I_c \propto \exp(-d/\xi_N)$ actually seems to favor the first alternative. Note that the density ρ only refers to the superconducting pairs. The total charge and current density comprises the density of the normal electrons. Local current conservation requires therefore that the normal electrons compensate locally for any unbalance in the superconducting densities. However, if we consider a situation with high-frequency currents (in which case we must also consider the term $\partial_t \rho$ for the pairs density and its normal counterpart $\partial_t \rho_N$) it may be impossible to obtain an exact local balance at all times.

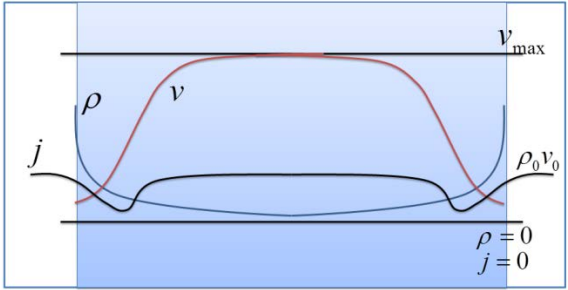

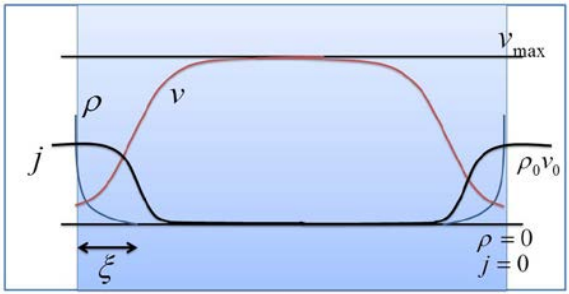

**Figure 2.** Possible behavior inside a thick inter-grain junction of the pairs density $\rho=|\Psi|^2$, the supercurrent density $j$, and the pairs velocity $v$, if the Ginzburg-Landau equation is only a poor approximation of the real non-local theory. Top: supposing that the increase in $v$ is slower than the decrease of $\rho$, but can catch up in the middle of the junction. Bottom: supposing that the increase in $v$ is insufficient to keep $j$ constant.

## 4. Conclusion

We are led to conclude that when the amplitude $\rho=|\Psi|^2$ decreases very sharply in the junction, one cannot suppose to write a locally-conserved microscopic expression for the current. Fig. 2 shows the qualitative behavior of the supercurrent density, under two different assumptions. In both cases the divergence $\frac{\partial j}{\partial x}$ is not zero in two regions of thickness $\xi_N$ inside the junction. It is not clear whether the normal current is able to compensate for this.

At the very least, this shows that the representation $j=\rho v$ is not adequate and therefore, if it is confirmed that $\partial_\mu j^\mu$ is not zero everywhere in certain states, because of a quantum anomaly, this should not be regarded as physically absurd in view of a classical local balance of charge ingoing/outgoing from a region.

A suitable formalism for the description of situations of this kind could be that of fractional quantum mechanics [17,18]. As recently shown by Wei [19], the probability current of the fractional Schrödinger equation is not in general locally conserved. Our covariant formulation of Aharonov-Bohm electrodynamics allows coupling of electromagnetism to charged fractional quantum systems [20]; this would be inconsistent in standard electrodynamics.

**APPENDIX: FURTHER REFERENCES**

*This brief review with an improved list of references has been kindly provided by L.M. Hively.*

Much work has been devoted to a more complete electrodynamic theory. Fock and Podolsky (1932) wrote the electrodynamic Lagrangian with the new term ($-C^2/2\mu$), but did not derive the resultant dynamical equations ($C=\partial_\mu A^\mu$). Ohmura (1956), Tomilin (2009), Tomilin (2013), Nefyodov and Smolskiy (2012), Gersten and Moalem (2015), and Alexeyeva (2016) included the gradient term ($\nabla C$) and the time-derivative term ($\partial C/\partial t$). Aharonov and Bohm (1963) included the new Lagrangian term ($-C^2/2\mu$) and derived the revised theory, including the Hamiltonian. Van Vlaenderen and Waser (2001) included the gradient term ($\nabla C$) and the time-derivative term ($\partial C/\partial t$), then derived the wave equation for $C$, as well as the revised forms for the momentum and energy balance. A fully hyperbolic formulation resolves charge non-conservation in electrodynamic simulations [Munz *et al.* 1999]. Arbab and Satti (2009) used a quaternion formulation, along with prediction of the SLW, which they call an electroscalar wave. Hively and Giakos (2012) used this revised formulation to predict a new free-space, scalar-longitudinal (or electro-scalar) wave with a scalar field ($C$) and an electric field along the propagation direction for **B**=0.

Jiménez and Maroto (2011) proposed a quantum electromagnetic field in an expanding universe that is incompatible with $C=0$ in a time-dependent geometry in curved space-time. The Jiménez and Maroto (JM) theory predicts a dynamical, propagating, scalar state ($C = \partial_\mu A^\mu$) together with a longitudinal electric field, plus transverse-photon dynamics. $C$ is excited gravitationally with a cosmological energy density whose value agrees with observations for inflation at the electroweak scale. The JM theory is compatible with all local gravity tests and is free from classical or quantum instabilities. Ratra and Peebles (1988) proposed a self-interacting, scalar field for dark energy. Woodside (2009) published the only work that derives the extended electrodynamics from first principles (4-vector geometry).